\documentclass{emulateapj}
\usepackage{rotating,psfig,natbib}

\begin{document}
\bibliographystyle{apj}

\newcommand{\Mearth}{M$_\oplus$}
\newcommand{\Msun}{M$_\odot$}
\newcommand{\Mjup}{M$_\mathrm{J}$}
\newcommand{\Rjup}{R$_\mathrm{J}$}
\newcommand{\Rearth}{R$_\oplus$}
\newcommand{\Rsun}{R$_\odot$}
\newcommand{\Lsun}{L$_\odot$}
\newcommand{\Rstar}{R$_\star$}
\newcommand{\Rlam}{$\mbox{R}_{\lambda}$}
\newcommand{\Teff}{T_{\rm eff}}
\newcommand{\AU}{\sc au}
\newcommand{\CsurO}{\rm C/O|_\odot}
\newcommand{\HOO}{H$_2$O}
\newcommand{\CHQ}{CH$_4$}
\newcommand{\COD}{CO$_2$}
\newcommand{\corot}{\emph{CoRoT}}
\newcommand{\kepler}{\emph{Kepler}}
\newcommand{\spitzer}{\emph{Spitzer}}
\newcommand{\microns}{$\mu$m}
\newcommand{\hd}{HD~189733b}
\newcommand{\etal}{et al.}
\newcommand{\deriv}{\mathrm{d}}
\newcommand{\ratiothree}{(R_p/R_\star)_{5.8\,\mu\mathrm{m}}}
\newcommand{\ratioone}{(R_p/R_\star)_{3.6\,\mu\mathrm{m}}}
\newcommand{\ratiotwo}{(R_p/R_\star)_{4.5\,\mu\mathrm{m}}}
\newcommand{\ratiofour}{(R_p/R_\star)_{8\,\mu\mathrm{m}}}

\title{Search for Carbon Monoxide in the atmosphere of the Transiting Exoplanet \hd\ }

\author{Jean-Michel~D\'esert\altaffilmark{1}, Alain~Lecavelier des Etangs\altaffilmark{1}, Guillaume H\'ebrard\altaffilmark{1}, David~K.~Sing\altaffilmark{1}, David Ehrenreich\altaffilmark{2}, Roger~Ferlet\altaffilmark{1}, and Alfred~Vidal-Madjar\altaffilmark{1}}

\altaffiltext{1}{Institut d'astrophysique de Paris, CNRS (UMR 7095),
Universit\'e Pierre et Marie Curie, 98 bis boulevard Arago, 75014
Paris, France}
\altaffiltext{2}{ Laboratoire d'astrophysique de l'observatoire de Grenoble, Universit\'e Joseph Fourier, CNRS (UMR 5571), BP53 38041 Grenoble cedex 9, France}

\begin{abstract}

Water, methane and carbon-monoxide are expected to be among the
most abundant molecules besides molecular hydrogen in the hot
atmosphere of close-in extrasolar giant planets. Atmospheric
models for these planets predict that the strongest
spectrophotometric features of those molecules are located at
wavelengths ranging from 1 to 10~\microns\, making this region of
particular interest. Consequently, transit observations in the
mid-IR allow the atmospheric content of transiting planets to be
determined. We present new primary transit observations of the
hot-jupiter HD\,189733b, obtained simultaneously at 4.5 and
8~\microns\ with the Infrared Array Camera (IRAC) onboard the
\spitzer\ \emph{Space Telescope}. Together with a new refined
analysis of previous observations at 3.6 and 5.8~\microns\ using
the same instrument, we are able to derive the system parameters,
including planet-to-star radius ratio, impact parameter, scale of
the system, and central time of the transit from fits of the
transit light curves at these four wavelengths. We measure the
four planet-to-star radius ratios, to be $\ratioone = 0.1545 \pm
0.0003$, $\ratiotwo = 0.1557 \pm 0.0003$, $\ratiothree = 0.1547
\pm 0.0005$, and $\ratiofour = 0.1544 \pm 0.0004$. The high
accuracy of the planet radii measurement allows the search for
atmospheric molecular absorbers. Contrary to a previous analysis
of the same dataset, our study is robust against systematics and
reveals that water vapor absorption at 5.8~\microns\ is not
detected in this photometric dataset. Furthermore, in the band
centered around 4.5\,$\mu$m we find a hint of excess absorption
with an apparent planetary radius $\Delta R_p/R_* =
$0.00128$\pm$0.00056 larger (2.3$\sigma$) than the one measured
simultaneously at 8\,$\mu$m. This value is 4$\sigma$ above what
would be expected for an atmosphere where water vapor is the only
absorbing species in the near infrared. This shows that an
additional species absorbing around 4.5~\microns\ could be present
in the atmosphere. Carbon monoxide (CO) being a strong absorber at
this wavelength is a possible candidate and this may suggest a
large CO/\HOO\ ratio between 5 and 60.

\end{abstract}


\section{Introduction}

\label{sec:intro}

The theory of transmission spectroscopy was developed in detail by
the pioneering works of Seager \& Sasselov (2000), Brown (2001),
Burrows et al. (2001), and Hubbard et al. (2001). This method is
now widely used to probe the atmospheric structure and composition
of the transiting planets. Applied to the first transiting
extrasolar planet known, HD\,209458b, this method allowed for the
first detection of an extrasolar planetary atmosphere (Charbonneau
\etal\ 2002), and of the discoveries of an escaping exosphere
(Vidal-Madjar \etal\ 2003, 2004, 2008; Ehrenreich et al. 2008),
and a hot hydrogen layer in the middle atmosphere (Ballester
\etal\ 2007).

HD\,189733b is a transiting hot-jupiter which orbits a small and
bright main sequence K star, and shows a transit occultation depth
of $\sim$2.5\% (Bouchy \etal\ 2005). The planet has a mass of $M_p
= 1.13$~Jupiter mass (\Mjup) and a radius of $R_p = 1.16$~Jupiter
radius in the visible (Bakos \etal\ 2006b; Winn \etal\ 2007). The
short period of the planet ($\sim$2.21858\ days) is known with
high accuracy (H\'ebrard \& Lecavelier des Etangs 2006; Knutson
\etal\ 2009). Sodium in the atmosphere of the planet has been
detected through ground-based observations (Redfield et al. 2008).
Using the Advanced Camera for Survey aboard the \emph{Hubble Space
Telescope} (HST), Pont et al. (2008) detected atmospheric haze
interpreted as Mie scattering by small particles (Lecavelier des
Etangs et al. 2008a).

The red part of the visible to the Infrared (IR) is a spectral
region of particular interest to detect molecular species using
transmission spectroscopy (e.g., Tinetti et al. 2007). The
rotational-vibrational transition bands of water (H$_2$O), carbon
monoxide (CO), and methane (\CHQ) are anticipated to be the
primary sources of non-continuum opacity in hot-Jovian planets.
Their relative abundances are determined by the C/O ratio (Liang
et al. 2003, 2004; Kuchner \& Seager 2006). Carbon monoxide is
expected to be the dominant carbon-bearing molecule at high
temperatures and \CHQ\ the dominant one below $\sim 1000$\,K
(Kuchner \& Seager 2006; Cooper \& Showman 2006; Sharp \& Burrows
2007). In the case of HD\,209458b's atmosphere,  Brown et al.
(2002) and  Deming et al. (2005) reported exploratory attempts to
detect  CO during transit from ground based observations. In the
same atmosphere, the possible presence of H$_2$O (Barman et al.
2005), TiO and VO (D\'esert et al. 2008; Sing et al. 2008a,b), and
the detection of the main constituent, H$_2$, through Rayleigh
scattering (Lecavelier des Etangs 2008b), have been reported using
observations in the visible from the Space Telescope Imaging
Spectrograph aboard HST.

Richardson \etal\ (2006) obtained the first infrared (IR) transit
measurement for this planet using the \spitzer\ \emph{Space
Telescope} (Werner et al. 2004) at 24~\microns, and derived
the planet radius with high accuracy.

\hd\ has been observed at 8~\microns\ (Channel~4) with the
Infrared Array Camera (IRAC; Fazio et al. 2004) aboard \spitzer.
The planet-to-star radius ratio at 8~\microns\ was found to be
$(R_p/R_\star)_\mathrm{8\,\mu m} = 0.1545 \pm 0.0002$ (Knutson
\etal\ 2007a). Agol et al. (2008) also observed this planet at
8~\microns\ during 7 primary transits and used the high
photometrical precision at this wavelength for the searching
companions using \emph{Transit Timing Variations} (TTV) method.

More recently, secondary eclipse measurements revealed the
presence of water, carbon monoxide and carbon dioxide in the
dayside emission spectrum of the planet, between 1.5 and
2.5~\microns\ (Swain \etal\ 2009), and water absorption with the
possible presence of carbon monoxide as a contributor near the
\spitzer/IRAC 4.5~\microns\ (Grillmair \etal\ 2008).

Results at 3.6~\microns\ and 5.8~\microns\ (Channel~1 and 3) from
primary transit observations of \hd\ using \spitzer/IRAC have
already been published but led to two different conclusions
(Ehrenreich et al. 2007; Beaulieu et al. 2008). From the analysis
of Beaulieu et al. (2008); Tinetti et al. (2007) concluded that
water vapour is detected in the atmosphere of the planet, whereas
Ehrenreich et al. (2007) claim that uncertainties on the
measurements are too large to draw any firm conclusions.
Independently, a detection of water and methane has also been
obtained using observations between 1.5 and 2.5~\microns\ from the
Near Infrared Camera and Multi-Object Spectrometer aboard HST
(Swain et al. 2008).

Here we describe \spitzer\ observations collected during two
primary transits of HD\,189733b, with the intent to measure its
radius at the four IR bandpasses of the IRAC instrument. We
analyze, in a consistent way, the observations from four bands
centered at 3.6, 4.5, 5.8 and 8 $\mu$m. We probe the planet
atmospheric composition by comparing results from these four
photometric bands. We first describe the observations and data
reduction in Sect.~\ref{sec:obs}. In Sect.~\ref{sec:ana}, we
describe the model and techniques used to estimate the physical
and orbital parameters of HD\,189733b. We discuss our results in
Sect.~\ref{sec:res} in the light of previous analysis and
theoretical predictions. We finally summarize our work in
Sect.~\ref{sec:sum}.

\section{Observations and data reduction}
\label{sec:obs}

\subsection{Observations}
We obtained \spitzer\ Guest Observer's time in Cycle 3 and 4 (PI:
A. Vidal-Madjar; program IDs 30590 \& 40732), in the 4
\spitzer/IRAC bandpasses. Our primary scientific objectives were
to detect the main gaseous constituents (\HOO\ and CO) of the
atmosphere of the hot-Jupiter \hd. First observations of the
system were performed on 2006 October 31 simultaneously at 3.6 and
5.8~\microns\ (channels 1 and 3) (Ehrenreich et al. 2007). The
second part of the program was completed on November 23, 2007
using the 4.5 and 8~\microns\ channels (channels 2 and 4)
following the same methods. The system was observed using IRAC's
stellar mode during 4.5 hours for each visit, upon which 1.8h was
spent in planetary transit. The observations were split in
consecutive subexposures at a cadence of 0.4 s for channels~1 and
2 and 2.0 s for channels 3 and 4. We obtained a total of 1936
frames for channels 1 and 3, and 1920 frames for channels 2 and 4.

While planning the observation in the Astronomical Observing
Request (AOR) format, we carefully selected a pixel area avoiding
dead pixels. We also purposely did not dither the pointing in
order to keep the source on a given pixel of the detector. Doing
so minimizes errors from imperfect flat-field corrections and thus
increase the photometric accuracy. This common observational
strategy matches that of recent \emph{Spitzer} observations of
HD~189733 and GJ 436 (Knutson et al. 2007; Deming et al. 2007,
Gillon et al. 2007; Agol et al. 2008)

\subsection{Data reduction}

We used the \spitzer/IRAC Basic Calibrated Data (BCD) frames in
the following analysis. These frames are produced by the standard
IRAC calibration pipeline and include corrections for dark
current, flat-fielding, detector nonlinearity, and conversion to
flux units. We first find the center of the PSF of the star to a
precision of $0.01$ pixel using the DAOPHOT-type Photometry
Procedures,  \texttt{CNTRD}, from the IDL Astronomy Library
\footnote{{\tt http://idlastro.gsfc.nasa.gov/homepage.html}},
which computes the centroid of a star using a derivative search.
We find that the position of the center of the star varied by only
less than $10\%$ of a pixel during the whole observation. We also
tested the position of the centroid of the star using
\texttt{GCNTRD}, which computes the stellar centroid by Gaussian
fitting. As a final test, we used a weighted-position sum of the
flux procedure in a $5 \times 5$ pixel box centered on the
approximate position of the star. These two last methods give very
similar results, which are in agreement within 0.02\% with the
positions extracted using \texttt{CNTRD}. We used the
\texttt{APER} routine, which we customized to perform a weighted
aperture photometry by summing the weighted background-subtracted
flux, on each pixel, within an aperture of a given radius (Horne
1986; Naylor 1998). We used a radius of $5$ pixels to minimize the
contribution of HD~189733B (Bakos et al. 2006a), the closest star
in the field of view. We checked that our results remain the same
when using a radius varying from $4$ to $6$ pixels. The background
level for each image was determined with \texttt{APER} by the
median value of the pixels inside an annulus, centered on the
star, with an inner and outer radii of $16$ and $18$ pixels,
respectively. This constant level is subtracted from each pixel of
a subexposure to create the background-subtracted image. The
background errors are dominated by statistical fluctuations. The
PSF, used for weighting, is estimated in each channel as the
median of the background-subtracted fluxes. The estimated error on
the weighted integrated flux is calculated as the square-root of
the photon-noise quadratic sum (Horne 1986; Naylor 1998). The four
raw weighted light curves obtained with this method are plotted in
the top panel of Fig.~\ref{fig:time_series}. After producing a
time series for each channel, we iteratively select and trim
outliers greater than $3 \sigma$ by comparing the measurements to
a transit light curve model. Doing so, we remove any remaining
points affected by transient hot pixels. We discarded 24, 26, 46
and 19 exposures in channel~1 to 4 respectively, which represent 1
to 2\% of the total number of data points.

\section{Fitting the \emph{Spitzer} Transit Light Curve}
\label{sec:ana}

\subsection{Out-of-Eclipse Baseline}
\label{sec:base}

Because of instrumental effects, the measured stellar flux
out-of-transit is not constant, but is seen to vary in time. To
correct for the instrumental effects, we define a baseline
function of time for each Channel. The baseline function is time-
and wavelength-dependent and is used to normalize a given time
series. We find that a linear function of time represents well the
data in channels~l and 2. On the other hand, in Si:As based
detectors (channels~3 and 4), the effective gain, and thus the
measured flux in individual pixels, drift non linearly over time.
This effect, called detector ramp, is well documented (Deming et
al. 2005, 2006; Harrington et al. 2007; Knutson et al. 2007). To
correct for the ramp and other baseline effects, we adopted a
non-linear function of time $F_{baseline} = A_0 + A_1 \times t
+A_2 \log(t-t_0) + A_3 \log^2(t-t_0)$, where $F_{baseline}$ is the
flux of the central star HD189733 without planetary transit and
$t_0$ was fixed to a time a few minutes before the first
observations. As a check, we also tested linear, polynomial and
exponential baseline functions but find that the logarithmic
baseline provides the best results for channels~3 and 4 (see also
Sect.~\ref{sec:res}). We find that baselines other than
logarithmic introduce significant systematic errors when
determining planetary parameters as seen in Fig.~\ref{fig:rmsCh3}.
We fit the detector correction coefficients simultaneously with
the transit-related parameters, allowing us to take into account
how changes in the correction coefficients may impact the transit
parameters.

\subsection{Pixel-phase effect correction}
\label{sec:pix}

Telescope jitter and intra-pixel sensitivity variations are also
responsible for fluctuations seen in the raw light curves
(Fig.~\ref{fig:time_series} upper panel), most severe in
channel~1. We find that the pixel-phase effect is negligible in
our dataset for the three channels~2, 3 and 4. A description of
this effect is given in the \spitzer/IRAC data handbook (Reach et
al.\ 2006, p.~50; Charbonneau et al. 2005). The method reported by
Morales-Calder\'on \etal\ (2006) and also applied by the two
previous analyses of the present dataset for channel~1 and 3
(Ehrenreich et al. 2007; Beaulieu et al. 2008) consists in
calculating a pixel-phase dependent correction by fitting the
light curve variations induced by this effect using a function of
one single parameter for the X and Y target positions. We find
that this single order function, described with only one
parameter, poorly corrects  the light curves for the pixel-phase
effect, and thus, systematics remains in the residuals
(Sect.~\ref{sec:res} and Fig.~\ref{fig:binresiduals}). We conclude
that this function (Reach et al.\ 2006) is not appropriate to high
precision photometry. To better decorrelate our signal from
pixel-phase variations, we adopted a different correction, based
on a quadratic function as described by Knutson \etal\ (2008) where we added a cross term ($K_5$),
$F_{corr}=F(1+K_1(x-x_0)+K_2(x-x_0)^2+K_3(y-y_0)+K_4(y-y_0)^2+K_5(x-x_0)(y-y_0))$,
where $x_0$ and $y_0$ are the integer pixel numbers containing the
source centroid and $F$ is the measured flux of the central star
HD~189733. The $K_i$ coefficients are the decorrelation
factors that have to be derived from fits to the the transit
lightcurve (TLC). A quadratic function of $X$ and $Y$, with five
parameters, improves the fit compared to a linear function, from a
$\chi^2$ of $2077$ to $1947$ for $n=1912$ degrees of freedom.
Introduction of the cross term ($K_5$) also slightly improved the
fit.
We find that adding higher-order terms to this equation does not
improve the fit for channel~1. As an additional test, we applied
this decorrelation procedure to one of the bright stars in the
field of view, resulting in a time series that showed no
significant deviations from a constant brightness.

The intra-pixel sensitivity is also expected to contaminate the
photometry in channel~2. However, we did not notice such a strong
effect as seen in Channel~1. We concluded that the central star
was localized on a part of the array which has a flat photometric
response (pixel reference: 147.20,198.25). This part of the
detector might be of particular interest for the warm \spitzer\
mission.

\subsection{Transit light curve model with limb darkening (LD) corrections}

We parameterized the transit light curve with 4 variables: the
planet-star radius ratio $R_p / R_\star$, the  orbital distance to
stellar radius ratio $a / R_\star$, the impact parameter $b$, and
the time of mid-transit $T_c$. We used the IDL transit routine
\texttt{OCCULTNL} developed by Mandel \& Agol (2002) for the
transit light curve model.

For each channel, we calculated a theoretical limb-darkening model
(Kurucz 1979) with $T_{\mathrm{eff}}= 5000$K, $\log g = 4.5$,
[Fe/H] solar, and fit this model\footnote{See {\tt
http://kurucz.harvard.edu/grids/}} to derive the 4 non-linear
limb-darkening coefficient defined by Claret (2000) and presented
in Table~\ref{tbl:LD}. We found that accounting for the effects of
limb-darkening decreased the resulting best-fit transit depth by
$0.5\sigma$ at 8.0~\microns\ to $1\sigma$ at 3.6~\microns. We
tested the robustness of our result for several limb-darkening
corrections. We derived planet-to-star radius ratios using linear
and quadratic limb darkening coefficients. In these cases the
resulting best-fit transit depth decreased by $0.4\sigma$ for
channel 1 and 2 and by $0.8\sigma$ for channel 3 and 4 compared to
a 4th degree polynomial limb-darkening corrections. Finally, we
modified the values of the non-linear limb-darkening coefficients
(Claret 2000) presented in Table~\ref{tbl:LD} by 10\% and found
that the planet-to-star radius ratios changed by less than
$0.5\sigma$.

\subsection{The fitting procedure}
\label{sec:fit}

We performed a least-squares fit to our unbinned data over the
whole parameter space ($R_p / R_\star$, $a / R_\star$, $b$, $T_c$,
$A_i$, $K_i$). In order to find the best-fit observables to the
data, we used the \texttt{MPFIT} package\footnote{{\tt
http://cow.physics.wisc.edu/$\sim$craigm/idl/idl.html}}, which
performs a Levenberg-Marquardt least-squares fit. We combined the
baseline function and the pixel-phase correction described above
with the transit light curve function which takes limb darkening
correction into account. We applied this method on the four
channels independently to calculate the four observables.

\subsection{Mean values and errors determination}
\label{sec:bootstrap}

We used a bootstrap method to determine the mean value, the
statistical and systematical errors for the measured parameters.
The possible presence of correlated noise in the light curve has
to be considered (Pont et al. 2006).

The bootstrap technique we used take into account both the red and
white-noise. To estimate the systematical errors due to intrapixel
sensitivity and baseline corrections, we randomly padded the
beginning of the transit light curve from different phase before
-0.03. Few thousand transit light curves with different baseline
duration are produced that way. Additionally, we randomly varied
each photometric measurement, within their estimated error bars,
following a normal distribution in order to derive the statistical
errors on the derived parameters. Totally, $4000$ synthetic
transit light curves were produced. We then fitted
(Sect.~\ref{sec:fit}) these transit light curves to derive a new
set of parameters and to extract their means and their
corresponding 1$\sigma$ statistical and systematical errors.

As an additional test for the errors, we measured the errors using
the Prayer Bead method (Moutou et al. 2004, Gillon et al. 2007).
In this case, the residuals of the initial fit are shifted
systematically and sequentially by one exposure, and then added to
the eclipse model before fitting. The purpose of this procedure is
to take into account the actual covariant noise level of the light
curve. Using this method, we found negligible red-noise after
corrections and obtained errorbars equivalent to the systematical
and statistical errors derived with the bootstrap method presented
above.

We examined the residuals from the best fit of each synthetic
transit light curve and for each channel independently. We
measured a RMS residual between $2.5$ and $3.3 \times 10^{-3}$ on
normalized flux for all exposures, depending on the channel (See
case of Channel~4 bottom panel Fig.~\ref{fig:rmsCh3}). The RMS
residuals are 20\% larger than the expected photon-noise, and
stays constant over the duration of the observation. We find that
the scatter of residuals in binned exposures decreases with bin
size as $N^{-1/2}$ for bins of up to $150$ points for all the
channels, as expected for photon-noise (See channel~1, bottom
curve in Fig.~\ref{fig:binresiduals}).

\section{Results and Discussion}
\label{sec:res}

The results of the independent fits are given in
Table~\ref{tbl:radii}. The most interesting parameters are the $4$
planet-to-star radius ratios which are discussed in this section.

\subsection{Comparisons with previous results}

The main difference between our analysis of channels~1 and 3 and
the previous studies (Ehrenreich et al. 2007, Beaulieu et al.
2008) resides in the pixel-phase decorrelation and baseline
functions adopted.

In channel~1, the flux measured is strongly correlated with the
position of the star on the detector array. We tested the
influence of the pixel-phase decorrelation function on channel~1
by comparing a correction function with one parameter, as used in
the previous studies (Ehrenreich et al. 2007; Beaulieu et al.
2008), to a 2nd degree function with 5 parameters as described in
Sect.~\ref{sec:pix}. For each corrected photometric time-series,
the red noise was estimated as described in Gillon et al. (2006),
by comparing the rms of the unbinned and binned residuals. In the
case of channel~1, the fit is largely improved by fitting 5
parameters compared to only one parameter for the X and Y star
position and only negligible systematics effects remain in the
residuals (Fig.~\ref{fig:binresiduals}). Contrary to the two
previous analysis, removing photometric points which have an
extreme pixel-phase value does not change our result, since they
are well corrected by the present decorrelation methods. We thus
kept all the photometric measurements.

The baselines are known to be inherently linear for channel~1 and
2 and logarithmic for channel~3 and 4 (Knutson et al. 2008).
However, we tested three different baselines functions: polynomial
of one, two and three degrees, an exponential with a polynomial,
and logarithmic with a linear function as describe in
Sect.~\ref{sec:base}. We tested the robustness of each of these
different baselines using the same test as described in Ehrenreich
et al. (2007), i.e. by dropping the first exposures from the
beginning of the observations (exposure with a phase smaller than
-0.03). We found that for channels~3 and 4, the best fits are
obtained when using a logarithmic baseline (see lower panel in
Fig.~\ref{fig:rmsCh3}). In the case of a linear, polynomial or
exponential baseline, the fitted parameters show large variations
with the number of data point removed (Fig.~\ref{fig:rmsCh3},
upper panel). The radius dramatically changes according to the
number of points removed when using a linear or a third degree
polynomial baseline functions indicating that systematics errors
remain in the corrected data. In the case of a logarithmic
baseline, the radius extracted does not depend on the number of
removed points. The logarithmic function is the only one which
allows the observable parameters to oscillate around the same
value independently of removed exposures. Only small systematics
still remain and they are included in the final error bar.
Consequently, our estimates of the error bars on the measured
parameters in channel~3 and 4 are conservative; they are larger by
about 30\% than the error bars due to the photon noise.

The present study shows that Ehrenreich et al.\ (2007) and
Beaulieu et al.\ (2008) did not apply a sufficient pixel-phase
decorrelation for channel~1. Consequently, systematics remained in
these two previous studies (Fig.~\ref{fig:binresiduals}).
Furthermore, none of these studies did use a logarithmic baseline
for channel~3. This explains why Ehrenreich et al.\ (2007)
obtained larger systematics error bars in this channel~3 when
exploring this issue. Our analysis also shows that by using linear
baselines, the result of Beaulieu et al.\ (2008) at 5.8~\microns\
is affected by large systematics which are not included in their
error bars estimations. Having understood the reasons of the
discrepancies, the results in the present paper can be considered
with most confidence.

The planet-to-star radius ratio at 8~\microns\ that we derive
(Fig.~\ref{fig:radii} and Table~\ref{tbl:radii}) is in agreement
with a previous measurement obtained with the same instrument and
same channel, but in subarray mode (Knutson \etal\ 2007a). The
error bars we obtain are slightly larger, since they also includes
systematic errors from the baseline (Fig.~\ref{fig:rmsCh3}).

\subsection{Discussion}
\label{sec:co}

We derived the fitted parameters and their error bars in a
consistent way. From our fits, we evaluated at each wavelength the
radius ratio $R_p/R_*$, the impact parameter $b$, the system scale
$a/R_\star$ and the central time of the transit
(Table~\ref{tbl:radii}).

The measured central time of the transit can be compared to the
expected transit time from known ephemeris. Several observations
of transits of \hd\ have been published recently with improved
ephemeris (Bakos et al. 2006b; Winn et al. 2007a; Knutson et al.
2007; Pont et al. 2007; Knutson et al. 2009). We find that the
central times that we obtained are in agreement within the error
bars for each set of simultaneous observation, between channels~1
and 3 for the first epoch and channels~2 and 4 for the second one.

Although data from different channels have been fitted
independently, our measurements of the impact parameter and of the
system scale both show the same behavior as a function of the
channel: they are consistent between channels observed
simultaneously, but disagree between channels observed at the two
different epochs. This strongly suggests that unrecognized
systematics remains between the observations obtained at two
epochs; these systematics could be of instrumental or more likely
of astrophysical origin (e.g, star spots). We therefore consider
that comparison of radius ratios should preferentially be made
between measurements taken at the same epochs.

The impact parameter and the system scale values can also be
compared with the most accurate measurements (Pont et al. 2007;
$b=0.671 \pm 0.008$ and $a/R_\star=8.92 \pm 0.009$), where the
degeneracy between primary radius and orbital inclination could be
lifted. Our results from our primary observations at 3.6~\microns\
and 5.8~\microns\ are in agreement with previous analysis of the
same data set (Ehrenreich et al.\ 2007; Beaulieu et al.\ 2008),
but are inconsistent at 3$\sigma$ level with the Pont et al.\
(2007) values. This suggests that the measurements in these two
channels may be affected by, for instance, stellar spots. The
measurements of the impact parameter and the system scale in
channels~2 and 4 data at 4.5~\microns\ and 8~\microns\ are in
agreement with the measurements by Pont et al.\ (2007). This
reinforces the confidence in our determination of the physical
parameters for channels~2 and 4.

The planet-to-star radius ratio measured at 3.6~\microns\ is
larger than the value extrapolated from simultaneous measurements
at 5.8~\microns\ assuming that only water molecules contribute to
the absorption in this channel. Therefore, other species must be
present in the atmosphere of the planet and absorb at this
wavelength. The possible presence of methane (CH$_4$) could also
contribute to the opacity at 3.6~\microns. Interestingly, the
planetary radius measured at this wavelength is in agreement with
the extrapolation of the measurements obtained with HST/ACS in the
visible (Pont et al. 2007, 2008) and assuming Rayleigh scattering
as needed to interpret the large variations of the planet radius
between 0.55 and 1.05~\microns\ (Lecavelier des Etangs et al.
2008b).

Comparing the planet-to-star radius ratio obtained simultaneously
in channels~2 and 4, we find a difference $(\Delta
R_p/R_\star)_{4.5-8~\mu  m} = 0.00128 \pm 0.00056$. This
corresponds to a $2.3\sigma$ signature of an excess absorption at
4.5~\microns\ relative to 8~\microns. The radius at 4.5~\microns\
is also $4\sigma$ above the expected value if only water molecules
were contributing to the absorption in this channel (see
Sect.~\ref{sec:co}). At this wavelength, another absorbent is
needed. The new observations at 4.5~\microns\ are not affected by
the systematics one faces in at 3.6~\microns , mainly the pixel
phase effect, since the central star was localized on a part of
the array which has a flat photometric response. Hence data
reduction and analysis of data taken in channel~2 at 4.5~\microns\
are straight forward compared to data in the 3~other channels. As
a consequence, the parameter extraction in this channel is the
most robust of the $4$ channels; conclusions drawn from this
channel can be considered with more confidence.

\subsection{Possible CO signature and large C/O ratio
scenario} \label{sec:co}

From Fig.~\ref{fig:radii} and Table~\ref{tbl:radii}, a possible
excess of absorption could be present in channel~2 around
4.5\,$\mu$m by comparison with channel~4. At this wavelength, all
main atmospheric constituents (H$_2$, H$_2$O, CH$_4$) have no
strong spectral features, except CO (Sharp \& Burrows 2007;
Fig~3). We therefore conclude that CO could be a candidate for
identification of this possible absorption signature.


Following Eq.~2 of Lecavelier des Etangs et al. (2008a), if
$\Delta R_p$ is the variation of the apparent planetary radius
between channels~2 and~4, we have
$$
\Delta R_p=H\ln \frac{\xi_2 \sigma_2}{\xi_4 \sigma_4},
$$
where $H$ is the atmospheric scale height, $\xi$ and $\sigma$ are
the abundance and cross-section of the main absorbent in channels
2 and 4, respectively. For a temperature $T$, $H$ is given by $ H
= k T/\mu g$, where $\mu$ is the mean mass of atmospheric
particles taken to be 2.3 times the mass of the proton, and $g$
the gravity. For \hd, we have $H/R_*$=0.00045 at 1500\,K. With
$\Delta R_p/R_*$=0.00128$\pm$0.00056, we find
$$
\log _{10} \frac{\xi_2 \sigma_2}{\xi_4 \sigma_4} = 1.245 \pm 0.540
\left(\frac{T}{1500\,{\rm K}}\right)^{-1} .
$$

The H$_2$O  absorption cross-section is smaller in channel~2,
$\sigma_{\rm H2O }(4.5~\mu m) = 2 \times
10^{-21}$\,cm$^2$/molecule, than in channel~4, $\sigma_{\rm
H2O}(8~\mu m) = 2 \times 10^{-20}$\,cm$^2$/molecule (Sharp \&
Burrows 2007). Therefore, assuming that H$_2$O is the main
absorber at 8~\microns, we find that the planet-to-star radius
ratio at 4.5~\microns\ should be $(R_p/R_\star)_{4.5~\mu m} =
0.1534 \pm 0.0005$. This value is 4$\sigma$ below our effective
measurement. Therefore, species other than H$_2$O should be
present. Because CO has a strong absorption signature at this
wavelength, it is a possible candidate for explaining the larger
planetary radius measured at 4.5~\microns.

CO has a similar cross section at 4.5~\microns\ as water at
8~\microns\ : $\sigma_{\rm H2O}(8 \mu m) = \sigma_{\rm CO}(4.5\mu
m) = 2 \times 10^{-20}$\,cm$^2$/molecule, therefore, the two
measurements in channels~2 and 4 lead to the conclusion that
$$\log _{10} \xi_{\rm CO} / \xi_{\rm H2O} =  1.245 \pm 0.540
\left(\frac{T}{1500\,{\rm K}}\right)^{-1} .$$

This corresponds to CO abundance of 5 to 60 times the H$_2$O
abundance. Note that the lower limit of the CO/\HOO\ ratio ($\sim
6$) is about $3.5\sigma$ higher than the highest ratio at which CO
absorption at 4.5~\microns\ are overcome by \HOO\ absorption
(CO/\HOO\ $=0.1$). In other words, at CO/\HOO\ ratio above $=0.1$,
CO is the main absorbent and can be detected at 4.5~\microns ; our
measurements show an absorption $3.5\sigma$ above this level.

A C/O ratio above the solar value could enhance, at  high
temperature, the CO abundance with respect to the abundance of
H$_2$O in the atmosphere of HD189733b (Tinetti et al. 2007).
Indeed, carbon-rich planetary environments do exist: a C/O ratio
much larger than solar has been measured with \emph{FUSE} and
\emph{HST} in the planetary system of $\beta$ Pictoris (Jolly et
al. 1998; Lecavelier des Etangs et al. 2001; Roberge et al. 2006).
Planets evolving in such environments could have a high C/O ratio
in their atmospheres. The large CO/\HOO\ ratio would appear at odd
with the detection of \CHQ\ claimed by Swain et al. (2008), though
the presence of CO is more consistent with the temperature above
1000\,K as estimated from primary and secondary transit
observations. Furthermore, simulations show that vigorous dynamics
caused by uneven heating of tidally locked planets can homogenize
the CO and \CHQ\ concentrations (Cooper \& Showman 2006). However,
recent 3D models (Showman et al.\ 2008) shows that the planet
could experience dramatic day/night temperature variations along
the planetary limb. As a consequence, the detection of both CO and
\CHQ\ molecules by primary transit observations could be
compatible if they are located in different limb regions.  \\

\section{Summary}
\label{sec:sum}

We have performed and analyzed new \spitzer/IRAC primary eclipse
observations of \hd\ at 4.5~\microns\ and 8~\microns\ and
reanalyzed observations at 3.6~\microns\ and 5.8~\microns\ in a
consistent way. Our analysis is more robust against systematics
than the previous ones (Ehrenreich et al. 2007, Beaulieu et al.
2008), especially regarding the pixel-phase effect and baseline
correction. Systematic error corrections can now be applied in all
channels.

As a consequence, contrary to what was claimed in a previous
study, we do not find the excess absorption at 5.8~\microns\
compared to 3.6~\microns\ that was initially found by Beaulieu et
al. (2008) and interpreted by water absorption by Tinetti et al.
(2007). Therefore, other species such as methane must be present
in the atmosphere of the planet and absorb at this wavelength.
Noteworthy, the measured radius at 3.6~\microns\ is compatible
with the radius extrapolated assuming Rayleigh scattering
absorption by small particles.

Interestingly in channel~2, the star is centered in a region of
the array which has a flat photometric response. Therefore, the
resulting measurements in this channel are robust.

Comparing  the planet-to-star radius ratios obtained
simultaneously at 4.5 and 8~\microns, we noticed a $2.3\sigma$
excess absorption at 4.5~\microns. Furthermore, the radius ratio
at 4.5~\microns\ stands $4\sigma$ above the expected value if only
water molecules were contributing to the absorption. If this
signature is confirmed, it could be interpreted by absorption by
CO molecules. In that case, we estimate a ratio of CO/\HOO\
abundances of 5 to 60, possibly indicating a high C/O ratio.

The Near Infrared Spectrograph (NIRSpec) aboard the future
\emph{James Webb Space Telescope} will enable medium-resolution
spectra over a wavelength range from 1 to 5~\microns\, and thus
allow for clear identification simultaneously and unambiguously of
\HOO, \CHQ\, CO and \COD.

\acknowledgements We thank the referee who greatly
contributed to improve the paper, as well as H.~Knutson and
Ph.~Nutzman for helpful discussion on data reduction and
analysis. D.K.S.\ is supported by CNES. This work is based on
observations made with the \emph{Spitzer Space Telescope}, which
is operated by the Jet Propulsion Laboratory, California Institute
of Technology under a contract with NASA. D.E. acknowledges
financial support from the French \emph{Agence Nationale pour la
Recherche} through ANR project NT05-4\_44463.

\end{document}